\documentstyle[preprint,eqsecnum,aps,prb]{revtex}

\begin{document}
\title{Direct observation of pinned/biased moments in magnetic superlattices}
\author{P. Padhan and W. Prellier\thanks{%
prellier@ensicaen.fr}}
\address{Laboratoire CRISMAT, CNRS\ UMR 6508, ENSICAEN,\\
6 Bd du Mar\'{e}chal Juin, F-14050 Caen Cedex, FRANCE.}
\date{\today}
\maketitle

\begin{abstract}
We report the {\it pinned/biased} {\it moment} in the superlattices
consisting of ferromagnetic (FM) SrRuO$_3$ and antiferromgnetic (AFM) SrMnO$%
_3$. This superlattice system shows anisotropy and oriented pinning/biasing
in the field-cooled (FC) hysteresis loop. The in-plane cooling-field
provides antiferromagnetic orientations while out-of-plane cooling-field
provides ferromagnetic orientations to the pinned/biased moments. The spacer
layer thickness, strength and orientation of magnetic field, cooling field,
and driving current influence the pinning strength. We propose that the
magnetic structure is a repetition of $AFM/Pin/FM$($Free$)/$Pin$ unit below
a critical field to explain its magnetic and transport properties. The
transport behavior is discussed using the spin-dependent conduction.
\end{abstract}

\smallskip \newpage

A biased magnetic field has been observed on cooling the FM-AFM system below
the Curie temperature ($T_C$) of the $FM$ through the Neel temperature ($T_N$%
) of the $AFM$ in presence of a magnetic field\cite{1,2,3,4,5}. It is
belived that the biased field is responsible for the shift of the hysteresis
loop along the field axis which has been observed in a wide variety of $%
FM-AFM$ systems, many of which do not exhibit a simple spin structure at the
interface to the $FM$ -AFM layers or materials. In general, a biased field
is established through field-cooling in the film plane where the magnetic
easy axis of soft ferromagnetic materials normally lies in plane. A shift in
hysteresis loop along the magnetization axis in addition to the shift along
the field axis is also observed\cite{5a}. The authors have explained the
shift in hysteresis loop along the magnetization axis by the pinned
uncompensated spin at the interfaces. Recently Maat {\it et al.}\cite{6}
have shown that exchange bias can also be observed for the magnetization
perpendicular to the film plane in $Co/Pt$ multilayers biased by $CoO$. They
investigated the biasing in various directions and found substantially more
within the sample plane, which they related to the anisotropy of the single-$%
q$ spin structure of the $CoO$. Several theoretical models have been
proposed to explain the origin of the biased field. Indeed, most of the
theoretical models assume a single domain state of the ferromagnetic layer
and focus on the domain structure of the $AFM$ layer for different types of
interfaces. However, despite the enormous research done in this field, this
effect is poorly understood.

Here, we report the direct observation of pinned/biased moments of the $%
SrRuO_3$ ($SRO$) layers by the $SrMnO_3$ ($SMO$) layers in the $SRO/SMO$
superlattices grown on ($001$)- $SrTiO_3$ ($STO$) substrates. To the best of
our knowledge, this observation has not been reported so far. The presence
of pinned/biased effect can be realized in the magnetic hysteresis loop with
field range below certain critical magnetic field (H$_P$). Various factors
such as the $SMO$ layer thickness, strength and orientation of the external
magnetic field and cooling field influenced the strength of the
pinned/biased moments of $SRO$, thus providing a way to control it.
Consequently, this presents the tantalizing possibility of controlling the
pinning of a $FM$ layer by the $AFM$ layer in an oxide multilayer, which is
a necessary step towards a better understanding and improvement of modern
magnetic devices.

The fabrication with optimized growth conditions and structural
characterizations of the superlattices have been reported elsewhere\cite{8}.
The superlattice structures were synthesized by repeating $15$ times the
bilayer comprising of $20$-($unit$ $cell$, $u.c.$) $SRO$ and $n$-($u.c.$) $%
SMO$, with $n$ taking integer values from $1$ to $20$. In all superlattices, 
$SRO$ is the bottom layer and the modulated structure was covered with $20$ $%
u.c.$ $SRO$ to keep the structure of the top $SMO$ layer stable. The samples
were characterized by resistivity ($\rho $) and magnetization ($M$)
measurements, in addition to x-ray diffraction and transmission electron
microscopy. Transport and magnetization measurements were performed at $10$ $%
K$ with magnetic field along the [$100$] and [$001$] directions of $STO$.
The samples were cooled to a desired temperature ($T$) from room temperature
in the absence of electric and magnetic field to perform zero-field-cooled ($%
ZFC$) measurements. The field-cooled (FC) measurements were always performed
with the same orientation of cooling field.

$SrRuO_3$\ is known as a metallic $FM$, with a Curie temperature ($T_C$) $%
\sim 160$ $K$ in its bulk form\cite{9}. Similar transport and magnetic
behaviors are observed in ($80$ $nm$)$SRO/STO$ with easy axis along [$001$]
direction of $STO$, consistent with Ref.10. The saturation field ($H_S$),
coercive field ($H_C$) and saturation magnetization ($M_S$) along the easy
axis of this film are $0.4$ $tesla$, $0.17$ $tesla$ and $1.46$ $\mu _B/Ru$,
respectively. Its $ZFC$ and $FC$ magnetic hysteresis loop ($M-H$) remain the
same at $10$ $K$ under $0.1$ $tesla$ cooling field ($H_{FC}$). The
current-in-plane magnetoresistance ($MR$) of this sample with magnetic field
along [$100$] and [$001$] directions of the $STO$ is negative although it is
hysteretic and higher when $H$ $\bot $ $I$. In contrast, $SrMnO_3$ is an $%
AFM $\ with a Neel temperature close to $260$ $K$\cite{11} and crystallizes
in a cubic structure when sandwiched between perovskite layers inside a
superlattice\cite{8}.

Fig. 1 shows the zero-field-cooled (ZFC) magnetization at 10 K at various
magnetic fields oriented along the in-plane and out-of-plane directions of
the substrate for the sample with 3 u.c. thick SMO layer. The easy axis of
SRO remains same in the superlattices. The in-plane magnetization of the
superlattice gradually increases as the magnetic field increases and becomes
larger than the calculated value (1.6 $\mu _B$/Ru), based on the only
contribution from SRO layer. This larger value of the in-plane magnetization
indicates that the SMO layer contribute to the net magnetization of the
superlattice at higher magnetic field. However, the out-of-plane hysteresis
loop shows a clear M$_S$ and H$_S$ with enhanced H$_C$. In order to
understand the strong anisotropic nature of the ferromagnetic layer in the
superlattice and their magnetotransport behavior below H$_C$, we have
measured the minor hysteresis loops of this superlattice in the field range
between the saturation field of SRO and the out-of-plane H$_C$ of the
superlattice with n = 3. The minor ZFC hysteresis loops in the field range
of $\pm $ 1 tesla are symmetric with respect to the origin (Fig. 2a and 2b)
for magnetic field along the [$100$] and [$001$] directions of $STO$. The
gradual increase in magnetization with the increase in magnetic field even
above the $H_S$ ($0.4$ $tesla$) of $SRO$ indicates that the spin-orbit
coupling is modified in $SRO$ layer and that the $MnO_6$ octahedra at the
interface influences the magnetic state of the $RuO_6$ octahedra\cite{12,13}%
. The $H_C$ of $SRO$ layer ($0.02$ $tesla$ along [$100$] and $0.17$ $tesla$
along [$001$]) is reduced in the superlattices ($0.001$ $tesla$ and $0.0027$ 
$tesla$ respectively). The magnetization of $SRO$ layer ($1.46$ $\mu _B/Ru$)
is decreased to $\approx $ $0.6$ $\mu _B/Ru$ in the superlattices. This
large suppression of the $FM$ state of the SRO layer in the superlattice
suggests that it is strongly influenced by the $G$-type $AFM$ state of the $%
SMO$ layer\cite{11}. In the case of $G$-type spin ordering, the ($00l$)
planes show the staggered pattern of spin arrangement, which is the source
of spin frustration at the compensated $SRO-SMO$ interfaces as well as the
spin canting in the $SRO$ layer in the vicinity of the interfaces\cite{14}.
Thus, due to the presence of $SMO$ layer, the spin canting/frustration in
the $SRO$ layer is reducing the effective $FM$ layer thickness of the SRO
layer in the $SRO/SMO$ superlattice. In others words, the effective
ferromagnetic $SRO$\ layer thickness is decreasing by the presence of a
canted/frustrated spin in the SRO layer close to the interface, which will
be detailed hereafter (Fig.5b).

In general, the magnetic interactions across the interface between the $FM$
and $AFM$ are known as exchange coupling ($EC$), with phenomenological
features such as an enhancement and an unidirectional anisotropy of $H_C$ 
\cite{1,2,3,4,5}. To study the exchange coupling at the $FM-AFM$ interfaces,
we have measured the $FC$ hysteresis loop of this sample. The $FC$
hysteresis loop of the superlattice with n = 3 for in-plane and out-of-plane
orientations of the magnetic field are shown in Fig. 2a and 2b respectively.
It shows several interesting features. First, the center of in-plane as well
as out-of-plane $FC$ hysteresis loop is shifted along the magnetization
axis. Second, the $FC$ hysteresis loops show a negligibly small change in
the values of $H_C$ compared to the $ZFC$ hysteresis loop. Third, the values
of the in-plane magnetization in the $FC$ hysteresis loop is lower (Fig. 2a)
while the out-of-plane magnetization in the $FC$ hysteresis loop is higher
(Fig. 2b) than its corresponding magnetization in the $ZFC$ hysteresis loop.
These features indicate that the spin configuration that was realized in the 
$ZFC$ state is modified in presence of cooling magnetic field. From the
observed $ZFC$ and $FC$ hysteresis loop one can conclude that the
canted/frustrated spins are aligned antiferromagnetically in the presence of
in-plane cooling-field and are aligned ferromagnetically for out-of-plane
cooling-field. So we define the oriented interfacial canted/frustrated spins
as the pinned/biased moments at the inerfaces. The in-plane pinned/biased
moment can be defined as $M_P^{//}$ $=$ $M_S$($0$)$-M_S$($H_{FC}$). Taking
into account the weak diamagnetic response of the substrate, the $M_S$ has
been extracted by extrapolating the linear part of the ($M-H$) curve to $H=0$%
. The value of $M_P^{//}$ when $H$ is antiparallel to $H_{FC}$ is larger by $%
\approx $ $0.302$ $\times $ $10^{-4}$ $emu$ (a factor of 0.3) compared to
the value of $M_P^{//}$ when $H$ parallel to $H_{FC}$. This indicates the
presence of moments at the interfaces which do not flip $180{{}^{\circ }}$
with the flipping of the magnetic field.\ So the canted/frustrated layer,
partially close to the $SRO-SMO$ interface, is pinned/biased along the
direction of the cooling magnetic field.\ In other words, {\it this is a
signature of uniaxial pinning/biasing of moments at the interfaces.} The
value of $M_P^{//}$ changes significantly at cooling fields below $\pm $ $%
0.03$ $tesla$ and remains constant for higher values of $H_{FC}$. Similarly,
the out-of-plane pinned/biased moment $M_P^{\perp }$ can be defined by
analogy to the bias field as $M_P^{\perp }=\frac{M_R^{+}+M_R^{-}}2$, where $%
M_R^{+}$ and $M_R^{-}$ are field-increasing and field-decreasing remanent
magnetization respectively. The same sign of the field for increasing and
decreasing $M_R$ (Fig. 2b) indicates the {\it uniaxial pinning/biasing of
moments}. The value of $M_P^{\perp }$ increases with $H_{FC}$ and changes
negligibly when $H_{FC}$ $>$ $0.1$ $tesla$. $M_P^{\perp }$ also depends on
the magnetic field that is applied and becomes zero when a magnetic field
larger than $1.5$ $tesla$ is applied. We have also measured the $M_P^{\perp }
$ for various superlattices and the results are given in Fig. 3. It
decreases as the $SMO$ layer thickness increases above $1$ $u.c.$, and
remains the same for $n$ $>$ $7$. Since $M_P^{\perp }$ varies with $SMO$
layer thickness, this indicates that the $EC$ at the interfaces is a
combination of the exchange coupling (J$_{exch}$) between $SRO$ layer and $%
SMO$ layer and the interlayer exchange coupling (J$_{int}$) between the $SRO$
layers. Note that for $SRO/SMO$ superlattice, the Neel temperature of $SMO$
layer is higher than the Curie temperature of $SRO$ layer. Since the
exchange coupling also depends on the thermal energy, the physical processes
responsible for the effective exchange coupling ($J_{eff}$ \symbol{126}$%
J_{Exch}$ + $J_{int}$) is expected to be different from the AFM/FM system
where $T_C$ $>$ $T_N$.

From the ZFC and FC\ magnetization measurements of the $SRO/SMO$\
superlattices, we have observed a strong anisotropy and pinned/biased
moments. To understand the effects of these magnetic behavior we have also
studied their electronic transport in presence of magnetic field below H$_P$%
. The $ZFC$ and $FC$ current-in-plane magnetoresistance for various magnetic
fields ($MR$ - $H$) in the range of magnetic field ($\pm $ $2$ $tesla$)
below $H_P$ of the sample with $n$ $=$ $3$ for field along [$100$] and [$001$%
] directions of $STO$ are shown in Fig. 4. The ZFC out-of-plane $MR$ (Fig.
4b) is negative as well as positive with hysteretic and asymmetric nature.
As the field sweep starts, the $MR$ increases and shows a sharp change from
positive to negative value at $+$ $H_{flip}$. On reverse sweep of $H$ to
zero from + $2$ $tesla$, the $MR$ decreases with a lower value than the $MR$
in the field-increasing branch. As $H$ increases in the negative direction,
the $MR$ becomes positive until the field is smaller than $-$ $H_{flip}$ and
at $-$ $H_{flip}$, the $MR$ becomes negative. The negative field decreasing
branch is similar but opposite to the reverse positive field sweep branch.
In presence of $H_{FC}$ the out-of-plane $MR$ (Fig. 4a and 4c) is negative
as well as positive, less hysteretic, more asymmetric and higher in
magnitude compared to the $ZFC$ $MR$. The $ZFC$ in-plane $MR$ (Fig. 4e) is
negative, non-hysteretic and symmetric with respect to origin. In presence
of $H_{FC}$, the ($MR-H$) curve becomes asymmetric (Fig. 4d and 4f). For a
field applied along the direction of $H_{FC}$ the in-plane $MR$ is larger
than the opposite direction field. Furthermore, the origin of the $FC$ ($%
MR-H $) shows a small shift towards the field antiparallel to the $H_{FC}$.
These phenomena are not the cumulative effect of the interfaces because of
the shortening of the top conducing layer. We attribute this asymmetric
nature of the field-cooled ($MR-H$) loop to the {\it uniaxial pinning/biasing%
} of moments observed in the $FC$ magnetic hysteresis loop.

The $SRO/SMO$ superlattices exhibit anisotropy with the orientation of
magnetic field to the sample in ($MR-H$) as well as ($M-H$) measurements.
The major contributions to this anisotropy behavior is from the strong
anisotropy of the $SMO$ layers and the additional periodicity of the
magnetic layer along the out-of-plane direction of the sample. The $ZFC$
hysteresis loop measured with $H$ $>$ $H_P$ for both orientations of $H$
indicates that the hard axis of $SMO$ is along [$001$] direction of $STO$
(Fig. 1). At a field much below $4$ $tesla$ but larger than the $H_S$ ($0.4$ 
$tesla$) of $SRO$, both $H_C$ and magnetization (at $1$ $tesla$) of the
superlattice is lower compared to the thin film of $SRO$ on $STO$ by $\sim $ 
$95$ $\%$ and $\sim $ $32$ $\%$ in-the-film-plane and $\sim $ $84$ $\%$ and $%
\sim $ $52$ $\%$ out-of-plane respectively. This suggests that the ideal $%
SRO/SMO$ magnetic structure (Fig. 5a) is lost as the sample is cooled down
to $10$ $K$ due to the strong anisotropy of $SMO$ layer, crystallographic
and/or magnetic reconstructions and relaxation at the interfaces\cite
{12,13,14}. We attribute the suppression of $H_C$ and magnetization to the
pinning/biasing of $SRO$ layer by the $SMO$ layer due to the strong exchange
coupling between them (at a field below $H_P$). At $H$ $<$ $H_P$ the
magnetization results partially from the part of the $SRO$ layer which
rotates coherently with the magnetic field. This part of the $SRO$ layer is
identified as the free layer. Using this picture, we can model the ideal
structure as a repetition of $AFM/$($pin$)$/FM$($Free$)/($pin$) unit (Fig.
5b). In the $ZFC$ state, the net magnetization of the pin layer is
negligible, i.e., antiferromagnetic orientation of the spin in the pin/bias
layers. But in the $FC$ state, the net magnetization of the pinned/biased
layer is lower by the same value as $M_P^{//}$ for in-plane $H_{FC}$, while
for out-of-plane $H_{FC}$ it is increasing to a finite value equal to $%
M_P^{\perp }$. Since $M_P^{\perp }$ is much larger than the $\frac{%
M_R^{+}-M_R^{-}}2$ on both $ZFC$ and $FC$ states, we conclude that the
volume of the free layer is smaller than the volume of pinned/biased layer.
Thus, the effective volume of the free layer depends on the $SMO$ layer
thickness, magnetic field and cooling field. Since the FC hysteresis loop of
the superlattices shifts along the magnetization axis, this effect in $%
SRO/SMO$ superlattices are seen at $0.1$ $tesla$ field low enough not to
saturate the $FM$ magnetization of $SRO$ ($H_S$ $=$ $0.4$ $tesla$), these
processes must occur in the $SRO$ layer. This is in contradiction with the
shifts in FC hysteresis loop along the magnetic field axis, in a magnetic
field high enough to saturate the $FM$ magnetization - where the
irreversible process occurs at the interfaces and in the $AFM$ \cite
{1,2,3,4,5}. In the range of $1$ $tesla$ ($<$ $H_P$) magnetic field, the
orientation of spin in $SMO$ layer is along the film-plane, for field along [%
$100$] and [$001$] directions of $STO$. Since the anisotropy axis of $SMO$
layer is fixed, the magnetic field along the easy axis of $SMO$ layer,
decreases the angle between the magnetization of $SRO$ layer and the easy
axis of $SMO$ layer, while their angular separation increases as the
magnetic field is rotated $90{{}^{\circ }}$. So, the in-plane $H_{FC}$ may
induce bilinear coupling of the spins of $SMO$ and $SRO$ at the interfaces,
while the out-of-plane $H_{FC}$ induces biquadratic coupling.

Transport processes in magnetic structures as spin-dependent tunneling\cite
{15} and scattering of spin-polarized carriers\cite{16} are influenced by
the spin-orientations of the pinned/biased layers and free layers of SRO.
The $FC$ magnetic field dependent $MR$ in this structure can be explained by
using the spin dependent scattering\cite{17} and the uniaxially pin/bias
spin. When the net moment in the pin and free layers are parallel, the
in-plane $MR$ is higher and out-of-plane $MR$ is negative; while the
antiparallel alignment of the net moments in the bias/pin and free layers
results in a lower in-plane $MR$ and a positive out-of-plane $MR$. This
correlation between the $FC$ out-of-plane magnetization and $MR$ with the
change in magnetic field is sketched in Fig. 5(c).

In summary, the magneto-transport properties of $SRO/SMO$ superlattices
deposited on $(001)-STO$ substrates were studied. Our data provide the
direct evidence for the {\it manifestation of the uniaxial pinned/biased
moments }in the $FM/AFM$ superlattice. The pinned/biased moments becomes
uniaxial as the superlattice is cooled in presence magnetic field. The
in-plane cooling field orients pinned/biased moments antiferromagnetically
while they orient ferromagnetically with the out-of-plane cooling field. The
electronic transport in these superlattices shows the evidence of spin
coupling of the mobile carriers to the interfacial pinned/biased layer. The
field dependent in-plane $MR$ is negative while the out-of-plane $MR$ is
negative as well as positive. We explain the magnetization and $MR$ by the
spin dependent scattering due to the relative orientation of the net
magnetization of the pinned/biased and free layers. Since progress towards
understanding and use of spin-electronic is growing rapidly, these results
should provide fundamentally new advances in both pure and applied sciences.

Acknowledgments:

We thank A. Fert, J.M.\ Triscone,\ B.\ Raveau,\ B.\ Mercey, A.\ Pautrat,\
A.\ Maignan and H.\ Eng for their helpful discussions.\ This work is
supported by the Centre Franco-Indien pour la Promotion de la Recherche
Avancee/Indo-French Centre for the Promotion of Advance Research
(CEFIPRA/IFCPAR) under Project N${{}^{\circ }}$2808-1.

\newpage Figures Captions:

Fig. 1: Isothermal ($10$ $K$) zero-field-cooled magnetization of the ($20$ $%
u.c.$) $SRO$/($3$ $u.c.$) $SMO$ superlattice at various fields oriented
along the [$100$] and [$001$] directions of the substrate, respectively.

Fig.2(a) and (b): Isothermal ($10$ $K$) zero-field-cooled and field-cooled
magnetization of the ($20$ $u.c.$) $SRO$/($3$ $u.c.$) $SMO$ superlattice at
various fields oriented along the [$100$] and [$001$] directions of the
substrate, respectively.

Fig. 3 Out-of-plane field-cooled biased/pinned moment ($M_P^{\perp }$) of
several superlattices at $10$ $K$.

Fig.4 Current-in-plane zero-field-cooled and field-cooled magnetoresistance $%
MR$ ($MR=\frac{R(H)-R(H=0)}{R(H)}$) of the ($20$ $u.c.$) $SRO$/($3$ $u.c.$) $%
SMO$ superlattice at various fields at $10$ $K$. Panels a, b and c show the $%
-$ $0.1$ $tesla$ $FC$, $ZFC$, and $0.1$ $tesla$ $FC$ magnetoresistance
respectively at various magnetic fields along the [$001$] direction of the
substrate. Panels d, e and f show the $-$ $0.1$ $tesla$ $FC$, $ZFC$, and $%
0.1 $ $tesla$ $FC$ magnetoresistance respectively at various magnetic fields
along the [$100$] direction of the substrate. The arrows indicate the
directions of the field sweep with the thicker arrow denoting the
commencement.

Fig. 5(a) and (b) Schematic view of the cross section of two interfaces of $%
SRO/SMO$ multilayer at room temperature and $10$ $K$, respectively. (c)
Schematic comparison of $FC$ magnetization and magnetoresistance measured
with magnetic field lower than the critical pinning field oriented along the
[$001$] direction of the substrate. In the rectangle box, thick and thin
arrows represent the relative orientations of the pinned and free layer net
moments, respectively.

\end{document}